# Quasiparticles with fractional charges in fractionally filled systems

Alexander N. Grigorenko

*Department of Physics and Astronomy, Manchester University, Oxford Road, Manchester M13 9PL, United Kingdom*

**Abstract:** We study systems that approach a state possessing discrete symmetry due to different degenerate realizations for the system. For concreteness, we consider fractionally filled systems where degeneracy comes from the presence of identical sub-lattices. We show that such systems possess a new type of quasiparticles with fractional charges, which we refer to as fractyons. We discuss static and dynamic properties of these quasiparticles.

Corresponding author: sasha@manchester.ac.uk

1. Systems with symmetries are interesting and important for physics. Continuous symmetries provide conservation laws through the Noether theorem while the presence of discrete symmetries can significantly modify properties of quasiparticles in the system or generate real space dislocations. Often, symmetry of a degenerate system is broken and only some of degenerate states survive. This leads to an order parameter distributed in real space[1]. Examples include a ferromagnetic transition with magnetization being a continuous order parameter, a Peierls' transition with a discrete order parameter describing the shift of atoms in 1D chain, superconductivity, superfluidity, and many others[2].

2. When symmetry is broken, we cannot use the standard statistical averaging. Indeed, this averaging would restore the symmetry of a system due to non-vanishing probability of transition between states with different values of an order parameter. For example, the standard statistical averaging would give zero magnetization even in the ferromagnetic state due to the fact that there exists extremely small but none-zero probability of changing the direction of magnetization of a system interacting with a thermal bath. As a result, the magnetization of a finite ferromagnetic object would average to zero under statistical averaging. However, it is well known that magnetization of a reasonably large ferromagnetic sample interacting with a thermal bath will be fixed at some direction for extremely long time.

3. To deal with the problem of averaging, scientists introduced quasi-averaging[3] by adding a symmetry-breaking term (proportional to a small parameter $\lambda$) to the Hamiltonian of a system and calculating averages in the order of doing the statistical averaging first (including the thermodynamical limit if necessary) and then taking the limit of $\lambda \to 0$. This turned out to be a powerful procedure that allowed one to describe a whole set of degenerate systems (ferromagnetic, antiferromagnetic, superconducting, superfluidic, etc.). In most cases, quasi-averaging can be reduced to calculating statistical averages with a "pinned" order parameter – the fixed direction of magnetization for ferromagnetic materials or a "chosen" degenerate state for the case of discrete symmetries. This does not preclude fluctuations near the fixed order parameter for systems with continuous symmetries. These fluctuations are often described with the help of quasiparticles (e.g., magnons for ferromagnetic materials). For systems with discrete symmetries, there exist a large energy barrier for transition between degenerate states (which depends on the size of a system) and as a result, an order parameter is firmly pinned for a large time for reasonably large systems.

4. A symmetry-breaking term could be different in different parts of a studied system. Hence, quasi-averaging could lead to different values of an order parameter in different points of space. This leads to a whole set of topological objects: domain walls, strings, vortices, skyrmions, dislocations, etc[4]. These objects are stabilised by topology as they are defined by behaviour of an order parameter at some larger distances (or peripheries) of an object. They exist in real space (not *k*-space!) and often have interesting and weird dynamics (e.g., magnetic domain wall dynamics[5]).

5. Here, we consider systems near a point of degeneracy that leads to discrete symmetry. We show that such systems can possess quasiparticles with fractional charges, which, for simplicity, we will refer to as fractyons. These quasiparticles exist in real space as complex arrangements of particles; they are connected to degeneracy of an order parameter and basic topology and require pinning of an order parameter in space. Hence, they are difficult to envisage using the standard statistical averaging.

6. To be specific, we discuss fractionally filled systems (FFS) where some particles fill a lattice at a fractional filling number. Figure 1 describes a typical scenario. Let us assume that we have a system consisting of interconnected sites populated by particles of some charge, see Fig. 1a. A charge of a particle under question, q, could be an electric charge (for a nanostructured metallic system) or a quantized vortex flux (for a nanostructured superconductor) or anything else – see discussion of some experimental realizations below. For simplicity, we assume that sizes of the dots for electrons are small enough to realise the Coulomb blockade[6] so that there is a big barrier to put two electrons to the same dot. As electrons repeal each other, they would produce a Coulomb crystal arrangement on the studied lattice. Analogously, for quantized superconductor vortices, we assume that there is a big barrier to have two pinned vortices on the same pinning site. Connections between sites can be realized as links shown as blue lines in Fig. 1a or links through an additional under-layer. For clarity, we remove the connection links from the following images. We assume that the number of charged particles populating the system depends on external conditions (e.g., gating voltage for the electrically charged particles or magnetic field for superconducting vortices). We introduce a filling factor, $f$, as a ratio of the number of charged particles in a system over the total number of sites. When the filling number $f$ is an exact fraction, we have an exact FFS. It is clear that exact FFS possess discrete degeneracy. Indeed, there exist several identical realizations of exact FFS which differ by quasi-averages (or different "pinned" sites). Fig. 1b,c shows this degeneracy for $f=1/2$.

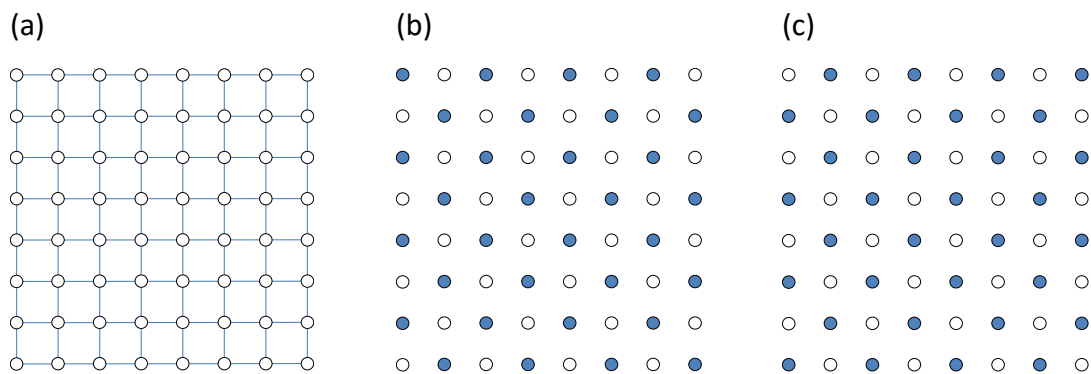

Fig. 1. Schematics of FFS. (a) A 2D square lattice for particle population. (b), (c) Two realizations of FFS at $f=1/2$. The blue sites are occupied by particles; the white ones are empty.

7. Let us additionally assume that we are at zero temperature. Then, at exactly $f=1/2$, only one of the degenerate states shown in Fig. 1b,c can be realized in the system due to the fact that fluctuations would not be able to overcome an energy barrier between these degenerate states. However, if the filling number is not exactly $f=1/2$ but close to it, then, the situation becomes more complicated. Let us add one more particle to FFS at $f=1/2$. One would expect that an additional particle should populate some empty site as shown in Fig. 2a.

Counterintuitively, it is not always the case and the minimum energy distribution of particles depends on particle-particle interaction. Indeed, other states with one additional particle are possible due to the presence of degeneracy at *f*=1/2, see Fig. 2b,c.

(a)

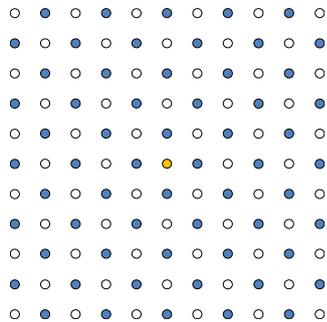

(b)

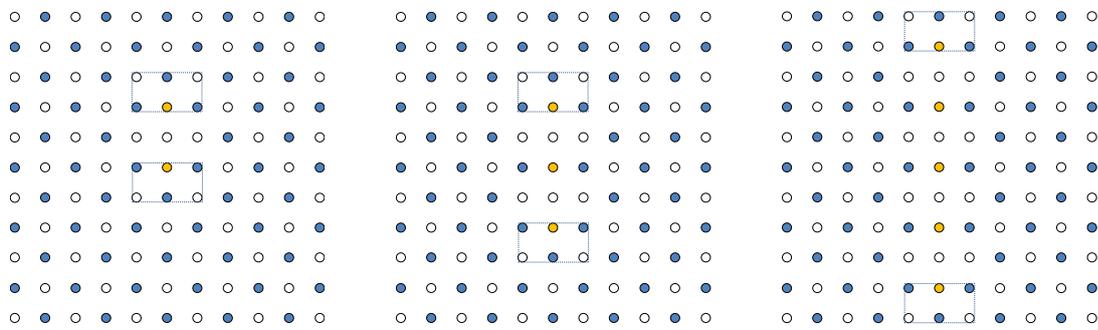

(c)

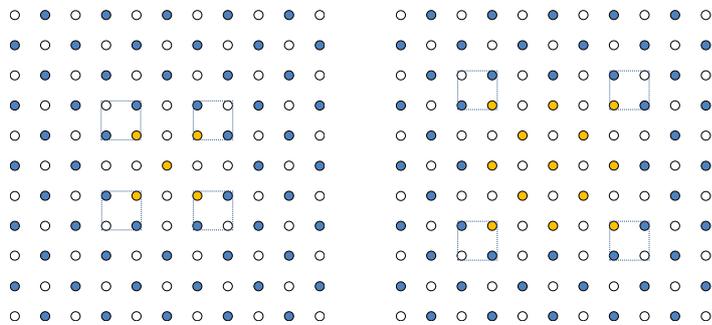

Fig. 2. Possible states with one additional particle added to FFS at *f*=1/2. (a) An "interstitial" state. Here, an additional particle (shown in yellow colour) populates a site of a sub-lattice that is degenerate to a "pinned" blue sub-lattice. (b) A "string" state. Blue dots show the pinned sub-lattice, yellow dots show the populated sites of the other degenerate sub-lattice. (c) A "square" state. Blue dots show the "pinned" sub-lattice, yellow dots show the populated sites of the other degenerate sub-lattice. Dotted rectangles (squares) show positions of fractyons described below.

It is worth stressing that all states depicted in Fig. 2 have just one additional particle added to *f*=1/2 state. The blue dots represent the original "pinned" FFS. The yellow dots represent particles occupying sites of an alternative degenerate sub-system. Figure 2b demonstrates "string" states, where the yellow particles produce a line while Fig. 2c shows "square" states, where the yellow particles (of an alternative sub-lattice) produce a square. One would expect that the energies of "string" and "square" states shown in Fig. 2b, c should be larger than the energy of an "interstitial" state shown in Fig. 2a due to the presence of domain walls (regions between blue and yellow populated sites). However, by introducing a string or a square of an alternative sub-lattice, we also reduce the interaction energy for an additional particle. Hence, the hierarchy of energies of states shown in Fig. 2 depends on particle-particle interaction.

8. To provide a simple physical picture for this hierarchy, let us take out an average charge associated with FFS $f$=1/2 from all the states. To do so, we average the charge over any four adjacent sites arranged in a square (which gives q/2 for FFS with $f$=1/2, where q is the charge of a particle that we use to populate the system) and then subtract it from the average charges for "interstitial", "string" and "square" states. Figures 3a-c show the effective states corresponding to those of Fig. 2a-c.

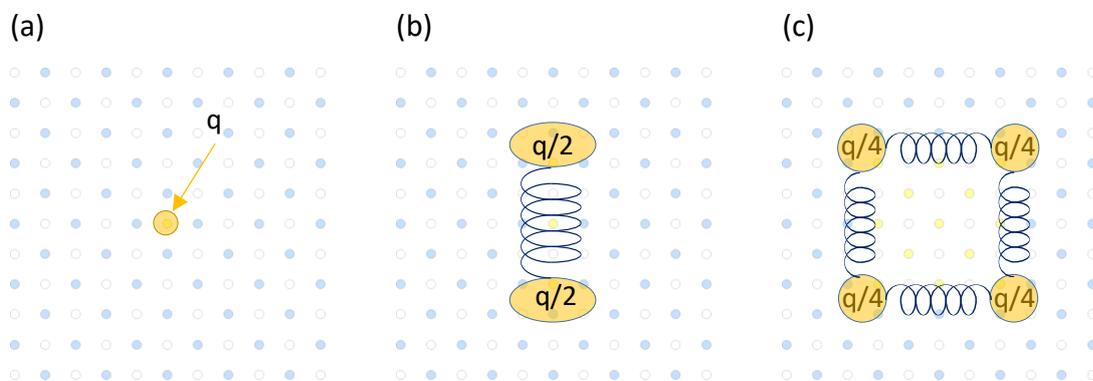

Figure 3. Effective states with one additional vortices after the background charge is subtracted. (a) An "interstitial" state with additional charge q. (b) A "string" state with two quasiparticles of charge q/2 connected by a spring. (c) A "square" state with four quasiparticles of charge q/4 connected by springs. The background transparent images show the initial images of Fig. 2.

As expected, an "interstitial" state corresponds to an additional particle of charge q. A "string" state corresponds to two quasiparticles of q/2-charge connected by a spring produced by a string of populated sites of an alternative sub-lattice (shown in yellow colour in Fig. 2). A "square" state corresponds to four quasiparticles with a charge of q/4 connected by springs produced at the perimeter of the domain of an alternative sub-lattice (shown in yellow colour in Fig. 2). We can see that the existence of an alternative degenerate sub-lattice produces the quasiparticles with fractional charges (q/2 and q/4, see Fig. 3b,c) which we will refer to as fractyons. This simple picture explains why energy of states with fractyons could be lower than energy of an "interstitial" state. Indeed, the energy of an additional charge introduced into the "pinned" sub-lattice (this energy is defined by particle-particle interaction energy) is proportional to the square of the charge. When we break an additional charge into $n$ quasiparticles with charge q/$n$ due to the degeneracy, the total energy is reduced by the factor of $n$. The presence of strings and domain edges between degenerate sub-lattices will add to this energy. If the reduction in energy due to charge splitting is larger than the increase of energy due to competing sub-lattices, then, the fractyons will be realized. Just for completeness, an effective energy for a "string" state with length $L$ can be written as $E_{string} = 2E_{self} - E_{rep}(L) + kL$. Here, $E_{self}$ is the energy of ½ fractyon, $-E_{rep}(L)$ is the repulsion energy of two ½ fractyons and $kL$ is the energy of the string. It is worth mentioning that the string energy is directly proportional to the length of the string (hence, an attraction force between fractyons is constant at large distances) while repulsion energy should have an asymptotic freedom (and disappear when distances between fractyons become comparable with the lattice constant). Historically, the presence of fractyons in superconducting vortices pinned by a square array at fractional number $f=1/2$ was first experimentally observed in[7]. It is clear that the longer the particle-particle interaction range, the larger the energy of an added particle would be and hence the larger the win of the energy connected to the charge splitting achieved in "string" and "square" states. Therefore, the "string" and "square" states should be energetically favourable for long-range particle-particle interactions.

9. To confirm the heuristic picture described in 8, we performed exact calculations of energies for the "string" and "square" states shown in Fig. 2. The energy of the system was calculated as $E = \frac{1}{2}\sum_{i \neq j} V(r_{ij})$, where $i$ and $j$ describe the sites occupied by particles and $V$ is the interaction energy between the particles. For simplicity, we choose an interaction energy between particles as $V(r) = \exp(-kr)$, where $k$ is the inverse of the interaction length and $r \geq a$ ($r$ is the distance between the particles and $a$ is lattice constant). In calculations, an area of >100x100 sites was studied – it was large enough to get rid of edge effects. Figure 4a shows a relative energy of the "string" states $E_{rel}(L) = E(L) - E(0)$ at one added particle as a function of the length of the string $L$ (measured in units of the lattice constant) for different values of the interaction length $k$. An "interstitial" state (that corresponds to $L=0$) was chosen as a baseline.

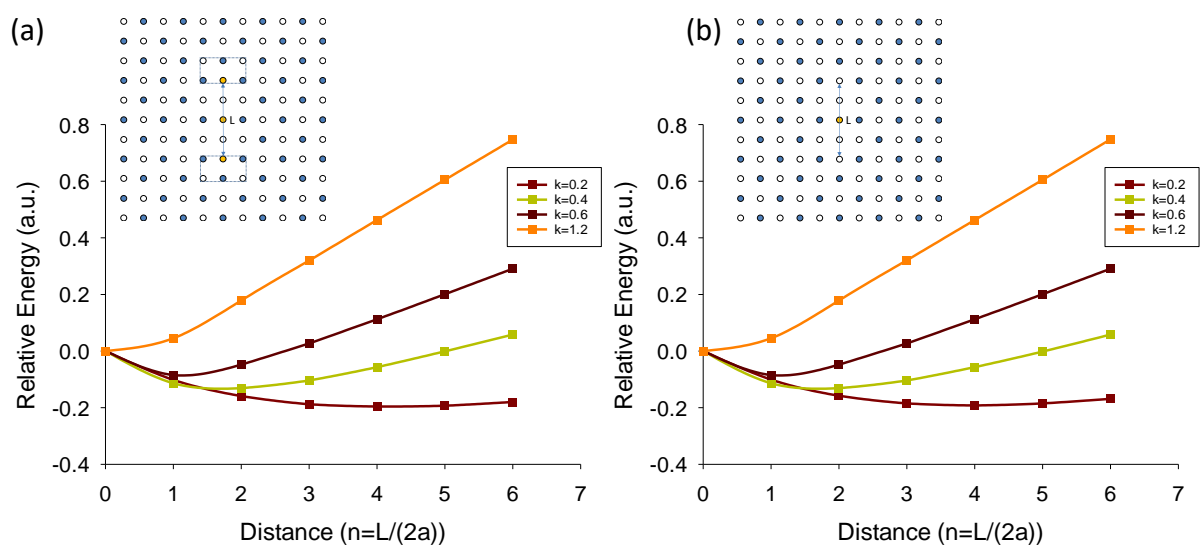

Figure 4. Relative energies $E_{rel}(L)= E(L) - E(0)$ of "string" states as a function of length of the string, $L$, at different values of interaction length described by $k$. (a) Energy of "string" states with an additional

particle. (b) Energy of "string" states with the absence of one particle ("hole" states). The insets show examples of string states for fractyons (a) and fractyon holes (b).

We see that for *k*=1.2 (which corresponds to a short interaction length compared to the lattice constant) the energy of an "interstitial" state is minimal as all relative energy $E_{rel}(L)$ are positive for any *L* > 0 and hence only "interstitial" states should be observed. With decrease of an interaction length (to *k*=0.6) a "string" state at *L*=2*a* becomes energetically more favourable than an "interstitial" state and corresponds to the energy minimum as a function of *L*. Further decrease of a particle-particle interaction length leads to minimum energy states with longer string lengths (*L*=4*a* for *k*=0.4 and *L*= 8a for *k*=0.2) in accordance with the heuristic theory of 8, see Fig. 4a. Calculations also show that "square" states are more energetically favourable than "string" states for small *k*. Table 1 provides a comparison of the minimum relative energy of "string" states and the energies of several "square" states evaluated at different values of *k*. From Table 1 we see that the "square" states are energetically favourable except for short interaction lengths where the "square" and "string" states are comparable in energy.

Table 1. Comparison of relative energies for "string" and "square" states. Energies are given in a.u. with respect to the energy of an interstitial state.

| k | $E_{rel\ string}$ | $L_{min}$ | $E_{rel\ square\ 2x2}$ | $E_{rel\ square\ 4x4}$ |
|---|---|---|---|---|
| 0.2 | -0.196 | 8*a* | -0.194 | -0.253 |
| 0.3 | -0.159 | 6*a* | -0.187 | -0.182 |
| 0.4 | -0.131 | 4*a* | -0.162 | -0.108 |
| 0.6 | -0.085 | 2*a* | -0.128 | -0.028 |
| 0.8 | -0.041 | 2*a* | -0.046 | 0.129 |
| 1.2 | 0.045 | 2*a* | 0.106 | 0.367 |

It is worth mentioning that "string" and "square" states can also be realised for a case of the absence of a particle from FFS at *f*=1/2. In this case, a "string" state would correspond to two fractyons of negative charge –*q*/2 connected by a spring (produced by a string of populated sites of an alternative sub-lattice). We will refer to these fractyons as fractyon holes. A "square" state in the absence of one particle corresponds to four fractyon holes of effective negative charge –*q*/4 connected by springs (produced at the perimeter of the domain of an alternative sub-lattice). The properties of fractyon holes mimic the properties of fractyons. Figure 4b shows relative energy of "string" states for fractyon holes as a function of the distance between the fractyons for different values of the particle-particle interaction length. We can see that the energy dependence (as a function of distance) for fractyon holes (Fig. 4b) almost repeats the energy dependence for fractyons (Fig. 4a). Hence, the "string" states become energetically favourable compared to single hole states at longer interaction lengths. As in case of fractyons, calculations show that "square" states will prevail at long interaction lengths for fractyon holes.

10. At zero temperature, the number of excess particles is defined by a mismatch between the filling factor $f$ and the exact fraction (e.g., for ½ FFS $N_{excess} = (f - 1/2)N$, where $N$ is the total number of sites.) At non-zero temperature, however, this is not the case as interplay between energy and entropy will introduce Frenkel defects[8] where some particles would vacate the "pinned" lattice to move to an alternative lattice in order to increase the entropy and hence to decrease the free energy of the system. As a result, at non-zero temperature, we will have some amount of interstitials and "hole" states even at exact filling factors due to Frenkel's arguments. These states will develop to either "string" or "square" fractyons states in order to reduce the total free energy. Concentration of Frenkel fractyons (and fractyon holes) can be evaluated as $n \approx N \exp\left(-\dfrac{\Delta E}{2k_B T_{fr}}\right)$ where $N$ is the number of sites, $\Delta E$ is the energy of a displacement of a "pinned" lattice particle to an alternative lattice, $k_B$ is the Boltzmann constant and $T_{fr}$ is the freezing temperature for the defects.

11. It is clear that the suggested phenomenon is quite general and can be observed in a variety of crystalline structures leading to quasiparticles (fractyons) of different charge fractions. There are two important cases to mention. The first case is connected to three-dimensional (3D) lattices. Here, fractyons can be fractured even more and an attraction between fractyons is realised through the membranes - domain walls between areas of different alternative lattices. For example, for a cubic lattice at $f=1/2$ discussed above, in addition to "string" and "square" states we will also have "cubic" states with fractyons of charge q/8 sitting at the corners of a cube and connected by membranes in addition to strings.

The second important case is connected to a hexagonal (triangular) lattice. This lattice epitomises complications arising for FFS at $f<1/2$. For simplicity, we consider a two-dimensional case of a hexagonal lattice at different filling factors of FFS. For $f=1/3$, we have three different arrangements of particles either of which can be "pinned". Figure 5a shows one possible realization. When we add one additional particle to FFS at $f=1/3$ we also can have three different degenerate realizations of the resulting state. Figure 5b shows one possible arrangement for this case. This implies that for $f=1/3$ we have more freedom and could observe fractyons of different charges. This is indeed the case. Figure 5c depicts a "trapezoid" state with two fractyons with charges of 2q/3 and one of –q/3. As in case of $f=1/2$, the fractyons appears after taking out the average charge for the FFS at $f=1/3$. All these charges are connected by springs that are produced by interfaces of different degenerate sub-lattices. It is worth stressing that the state shown in Fig. 5c still corresponds to just one additional particle added to FFS at $f=1/3$. When two particles are added to FFS at $f=1/3$, we could have a "triangular" state with three fractyons of a 2q/3 charge, see Fig. 5d. At the same time, for a hexagonal lattice at the filling factor of $f=1/4$ we can still have "string" states (with fractyons of q/2 charges) and "square" states (with fractyons of q/4 charges) shown in Fig. 5e,f which are analogous to "string" and "square" states of FFS at $f=1/2$ described above.

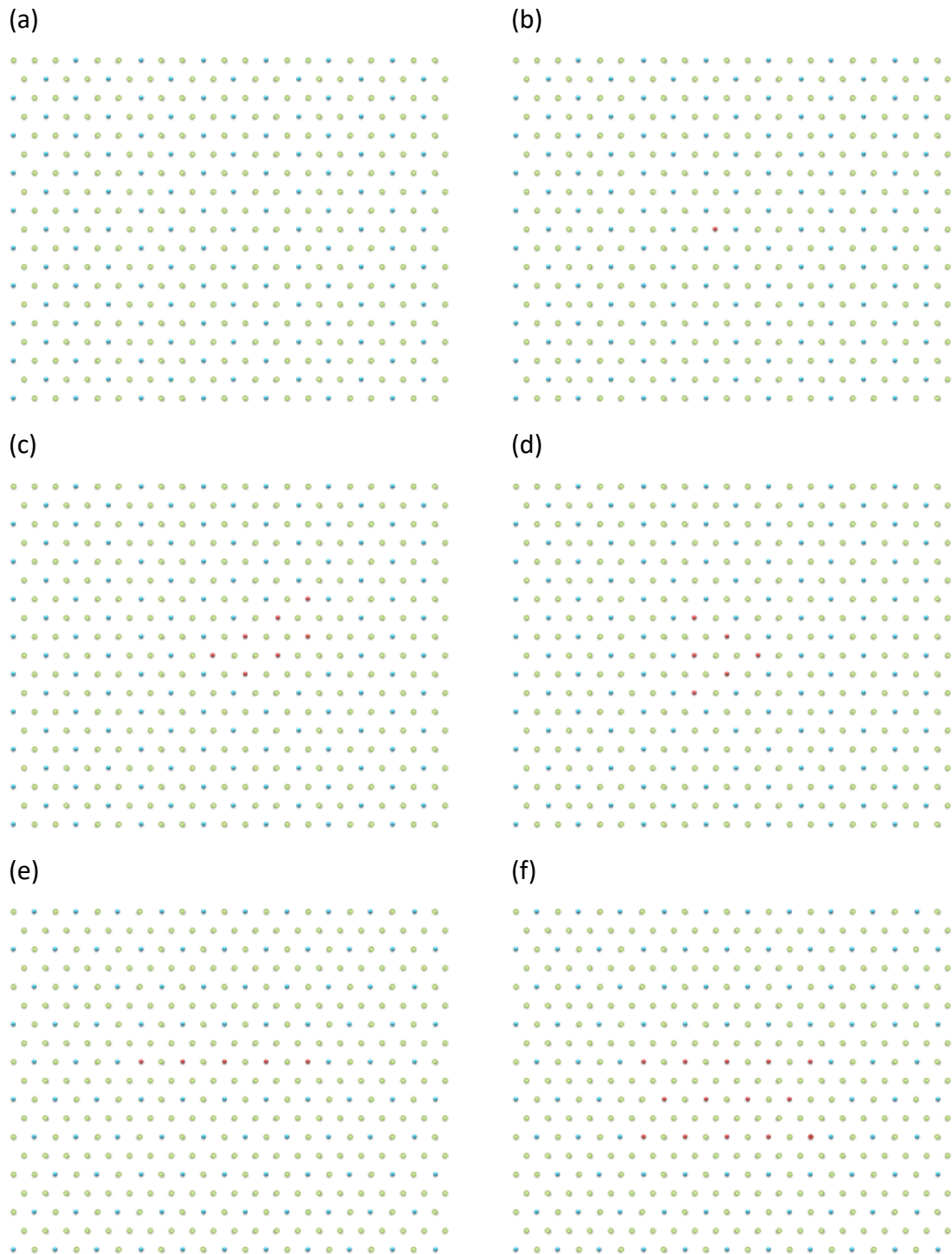

Figure 5. FFS states for a hexagonal lattice. The blue dots show the populated sites, the empty green dots show the vacant sites, the red dots show the populated sites of an alternative sub-lattice. (a) A hexagonal lattice at FFS with $f=1/3$. (b) A hexagonal lattice at FFS $f=1/3$ with one added particle shown in the red colour. (c) A "trapezoid" state for FFS $f=1/3$ with one added particle. The corners of the red site polygon have effective charges of $2q/3$, $2q/3$ and $-q/3$. (d) A "triangular" state with two added particles for FFS $f=1/3$. The corners of the triangle carry the charge of $2q/3$. (e) A "string" state with one added particle for FFS $f=1/4$ with two fractyons of charge $q/2$. (f) A "rectangular" state with one added particle and four fractyons of charge $q/4$.

12. The dynamics of fractyons is complicated. In the simplest 2D scenario, it can be described by fractional charge quasiparticles with some effective masses connected by springs as shown in Fig. 3b, c. The effective Hamiltonian of the system can be written as $H_{eff} = \frac{P^2}{2m_{eff}} + V_{eff}(L)$, where $P = p - \frac{f_q e}{c} A$ is the effective momentum of a fractyon, $m_{eff}$ is the effective mass of a fractyon and $V_{eff}(L)$ is the effective energy of fractyon interactions. Hence, one would expect to observe a natural vibration frequency of $\omega = \sqrt{\frac{K}{m_{eff}}}$ for spring-connected fractyons, where $K = \frac{\partial^2 V_{eff}}{\partial L^2}$. It is worth noting that these natural vibrations are dark and cannot be excited by external electromagnetic fields. Additional complications come from the fact that the effective mass tensor for fractyons (as well as spring constants) could be anisotropic depending on the lattice involved.

13. As we mentioned above, realizations of FFS and states close to it were first experimentally observed for superconducting vortices by scanning Hall probe microscopy in thin superconductor films with underlying pinning array[7]. Magnetic field (applied to a superconductor film with a pinning array for superconductor vortices) generated FFS at some values of magnetic fields. Another system where FFS could be observed is Coulomb crystals. Figure 6 shows a possible experimental realization for an artificial Coulomb crystal. Here an array of quantum dots (shown in yellow colour) with a Coulomb blockade (which requires reasonably small dot sizes) are placed on an ultrathin conducting layer (shown in green colour) separated from a conducting blue sublayer (say, silicon) by a thin layer of dielectric (say, silica). By applying a gate voltage to the bottom and the top layers, we introduce some number of electrons to the lattice that could result in FFSs.

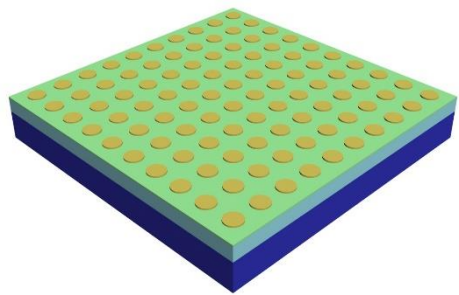

Figure 6. A possible experimental realization of FFS states in artificial crystals. The yellow dots represent metallic quantum dots; a thin green metallic sublayer is used for electrical connection between the dots; a light blue dielectric layer and a bottom blue conducting layer can be used for gating of the system.

14. It is difficult not to notice a resemblance of the studied FFSs with systems in which a fractional quantum Hall effect (FQHE) is observed[9]. Indeed, FQHE is observed at fractional filling of a Landau level, which is in the realm of the proposed theory. There is a mapping of properties of a single-layered superconductor in the London region to those of 2D electrons in linear approximation. Indeed, if we choose z-axis perpendicular to the sample surface, then z-component of magnetic field $h_z = \mathbf{n} \cdot \mathbf{H}$ (where $\mathbf{n}$ is the unit vector normal to the surface) for a 2D superconducting vortex (pancake vortex) will satisfy the following equation (in the London approximation)[10]

$$h_z - \lambda_L^2 \Delta h_z \delta(z) = \phi_0 \delta(\mathbf{r}), \qquad (1)$$

where $\lambda_L$ is the London penetration depth and $\phi_0 = \dfrac{2\pi\hbar c}{e}$ is the flux quantum. At the same time, electric potential $\varphi$ of an additional electron moving in 2D plane will satisfy an analogous equation (in the linear approximation)[11]

$$\varphi - \lambda_D^2 \Delta \varphi \delta(z) = \dfrac{4\pi\lambda_D^2 e}{\bar{\varepsilon}} \delta(\mathbf{r}), \qquad (2)$$

where $\lambda_D$ is the Debye screening length and $\bar{\varepsilon}$ is the average permittivity. We can see that (2) can be easily recast in the form (1). In addition, the energy interaction between monopoles $\phi_0$ has the same form as that between electrons. Hence, the properties of 2D electron gas should mimic well-studied properties of superconducting vortices in thin films. This implies that 2D electrons should form a Wigner crystal (a Coulomb crystal) at some electron density. At higher electron density, this crystal will melt due to the Lindeman criterion. At low electron density, the Wigner crystal of electrons could experience a re-entrant melting. This is indeed what theoreticians predicted for 2D electrons in magnetic field: Wigner crystals produced by 2D electrons are melted when the lowest Landau level is completely occupied but they can still exist at low filling fractions $f \leq 1/7$, see, e.g., Ref. [12]. At the same time, there is an experimental evidence[13] that Wigner crystals for 2D electrons in magnetic field can survive up to the fraction $f = 1/3$ of population of the lowest Landau level. When 2D electrons form a Wigner crystal at fractional filling factors in the quantum Hall regime, we can apply theory presented here and realise fractyons in a system much the same as they were realised in superconductor systems[7, 14]. The importance of fractyons for FQHE (if any!) is an interesting topic for investigations.

The properties of fractyons in a hexagonal lattice are also striking: they have charges of $\pm 2/3q$ and $\pm 1/3q$, these charges are connected by springs (which leads to fractyon confinement); while fractyons also show an asymptotic freedom. All these properties have uncanny resemblance to those of quarks. It could be just a coincidence or there might be a connection of SU(3) symmetry of strong interactions to discrete degeneracy of an appropriate fractionally field lattice where gluons represent strings that connect fractyons.

In conclusion, we have shown that fractionally filled systems with degenerate discrete states possess fractionally charged quasiparticles referred to as fractyons. These quasiparticles are generated by degeneracy of FFS states and cannot be observed in standard statistical averaging, as they require "pinning" of FFS states. They could be useful in various areas of science.

**Acknowledgement**

The author thanks Prof. Simon Bending for stimulating discussions and ideas.